\documentclass[useAMS,referee,usenatbib]{biom}
\usepackage{multirow}
\usepackage{lscape}
\usepackage[utf8]{inputenc}
\usepackage{graphicx}
\usepackage{tabularx}
\usepackage[english]{babel}
\usepackage{url}
\usepackage{parskip}
\usepackage{dsfont}
\usepackage{amsmath}
\usepackage{amssymb}
\usepackage{enumitem}  
\usepackage{mathtools}
\usepackage{mathrsfs}
\usepackage[autostyle]{csquotes}
\usepackage{algorithm,algpseudocode}  
\usepackage[normalem]{ulem} 
\usepackage{calc}
\usepackage{setspace}
\usepackage{hyperref} 
\usepackage{xcolor}
\usepackage{verbatim}

\def\bSig\bmath{\Sigma}

\newlength{\remaining}

\DeclareMathOperator*{\argmax}{arg\,max} %
\DeclareMathOperator*{\argmin}{arg\,min}

\newcommand\bstrut{\rule[-1.0ex]{0pt}{0pt}}

\title[Clustered ${Q}$-Learning for Examine Moderators in Constructing cAI]{$\bm Q$-Learning with Clustered-SMART (cSMART) Data: \\ Examining Moderators in the Construction of Clustered Adaptive Interventions}

\author{
Yao Song$^{1,\dagger}$, 
Kelly Speth$^{1,\dagger}$, 
Amy Kilbourne$^{2, 3}$, 
Andrew Quanbeck$^{4}$, Daniel Almirall$^{5, 6}$,Lu Wang$^{1, *}$
\email{luwang@umich.edu} \\
$^{1}$Department of Biostatistics, School of Public Health, University of Michigan, Ann Arbor, USA \\
$^{2}$Department of Learning Health Sciences, University of Michigan, Ann Arbor, MI, USA \\
$^{3}$Office of Research and Development, U.S. Department of Veterans Affairs\\
$^{4}$ Department of Family Medicine and Community Health, University of Wisconsin, Madison, WI, USA \\
$^{5}$Department of Statistics, University of Michigan, Ann Arbor, MI, USA \\
$^{6}$Survey Research Center, Institute for Social Research, Ann Arbor, MI, USA\\
$\dagger$: \text{These authors contributed equally to this work and are ordered alphabetically by last name}
}

\begin{document}


\date{{\it Received ???} 2025. {\it Revised 000} 0000.  {\it Accepted 000} 0000.}



\volume{0}
\pubyear{0000}
\artmonth{000}




\begin{abstract}
    A clustered adaptive intervention (cAI) is a prespecified sequence of decision rules that guides practitioners on how best - and based on which measures — to tailor cluster-level intervention to improve outcomes at the level of individuals within the clusters. A clustered sequential multiple assignment randomized trial (cSMART) is a type of trial that is used to inform the empirical development of a cAI. The most common type of secondary aim in a cSMART focuses on assessing causal effect moderation by candidate tailoring variables. 
    We introduce a clustered $Q$-learning framework with the $M$-out-of-$N$ Cluster Bootstrap using data from a cSMART to evaluate whether a set of candidate tailoring variables may be useful in defining an optimal cAI.  
    This approach could construct confidence intervals (CI) with near-nominal coverage to assess parameters indexing the causal effect moderation function. Specifically, it allows reliable inferences concerning the utility of candidate tailoring variables in constructing a cAI that maximizes a mean end-of-study outcome even when ``non-regularity", a well-known challenge exists.
    Simulations demostrate the numerical performance of the proposed method
    across varying non-regularity conditions and investigate the impact of varying number of clusters and intra-cluster correlation coefficient on CI coverage. 
    Methods are applied on ADEPT dataset to inform the construction of a clinic-level cAI for improving evidence-based practice in treating mood disorders.
\end{abstract}

%



\maketitle
%
\section{Introduction}
\label{s:intro}
In many health and education settings, intervention occurs at the level of a cluster \cite[e.g., clinic or school; ][]{raudenbush2020randomized}, with the intent of improving outcomes at the level of units within the clusters. 

A clustered adaptive intervention (cAI) is a pre-specified sequence of decision rules that guides practitioners on how best—and based on which measures—to tailor intervention options at the level of a cluster (e.g., clinic), with the expressed intent of improving one or more outcomes at the level of units within clusters (e.g., patients).
In a cAI, the provision of cluster-level intervention  occurs at a pre-specified set of decision points. At each decision point, intervention can be  tailored based on any observed measure(s) of change at the cluster-level, intervention options offered in prior stages, or their combination. These measures, which are known as \emph{tailoring variables,} can be impacted by intervention in prior stages. 

A clustered, sequential, multiple-assignment, randomized trial  \cite[cSMART; ][]{nahum2019} is a type of multi-stage trial in which clusters are randomized at two or more decision points in a cAI, and the primary outcome is at the level of units nested within each cluster (e.g., individuals within clinics).  
cSMARTs are designed  to answer scientific questions that aide the construction of effective cAIs \cite[][]{AlmirallNahum2018,nahum2012q}. 
The majority of applications to-date have focused on constructing adaptive implementation strategies. This is a special type of cAI where the goal is to improve the adoption,  implementation, or delivery of a known intervention by individuals (e.g., clinicians) within clusters (e.g., clinics) who deliver the intervention. For example, the goal of the ADEPT study \cite[Adaptive Implementation of Effective Programs Trial; ][]{kilbourne2014protocol} is to construct a cAI to improve the implementation of the Life Goals Intervention in community-based, primary care clinics; the goal of the ASIC study \cite[Adaptive School-based Implementation of CBT; ][]{kilbourne2018adaptive} is to implement Cognitive Behavioral Therapy in high-schools; and the goal of the BOI study \cite[Balanced Opioid Initiative; ][]{quanbeck2020balanced} is to improve guideline-concordant opioid prescribing (as recommended by the US Centers for Disease Control) in non-cancer primary care clinics.
In one application \cite[][]{kasari2021getting}, the focus is to construct a multilevel adaptive intervention---a type of cAI with multiple, nested levels of intervention---for children with autism.

A common primary aim in a cSMART concerns the comparison of two (or more) cAIs on the mean outcome at the individual-level. 
Existing methods for such primary aims include marginal mean models and associated inverse probability weighting methods \cite[][]{necamp2017comparing}, sample size formulae  \cite[][]{ghosh2015chapter,necamp2017comparing} and, more recently, finite-sample adjustment methods for making improved statistical inference in marginal mean comparisons \cite[][]{pan2024finite}.

This manuscript develops a statistical method to assess the utility of candidate tailoring variable(s) in an optimal cAI (one that maximizes a mean outcome) by way of effect moderator analyses. Using the method requires pre-specifying a set of cluster-level candidate tailoring variables at each decision point. The goal is to estimate whether (or the extent to which) the causal effect of one intervention option vs another varies across levels of the candidate tailoring variables (or some combination of them) at all decision points.  
Such aims are central to the justification of all SMARTs; yet, in the case of cSMARTs, the development of associated methods has received little-to-no attention, thereby limiting domain scientists/analysts from building their cAI science. 

For data from standard (non-clustered) SMARTs, the $Q$-learning regression methods in \cite{chakraborty2010inference}, \cite{chakraborty2013inference} and \cite{laber2014dynamic} were the first to effectively 
construct confidence intervals that deliver nominal coverage.  
These methods overcome an important technical challenge known as non-regularity \cite[][]{laber2011adaptive,chakraborty2013inference}, 
which arises because inference about causal effect moderation at all, but the last, decision points is based on non-smooth functions of subsequent-stage regression estimators. The non-smoothness is the result of setting future decision rules to their estimated optimum.
For cSMART data, these methods need to be extended to account for the nesting of units within clusters, and their performance evaluated. 
Thus, a key contribution of this manuscript is the development and evaluation of an $M$-out-of-$N$ Cluster Bootstrap sampling procedure to construct appropriate confidence intervals.  

We evaluate the method via simulation experiments that study the impact of varying total study sample size, effect size, and magnitude of the true intra-cluster correlation on the performance of the confidence intervals in terms of coverage rates and the statistical efficiency. We also provide freely-available R code to implement the proposed method. 
To illustrate the method, we use data from ADEPT \cite[][]{kilbourne2014protocol}, which is---to our knowledge---the first cSMART to be designed and conducted.


\section{Notation and Set-up}

\subsection{Observed Data}

We assume an observed data arise from a cSMART with $N$ clusters, indexed by $i=1,\ldots,N$. Within cluster~$i$, there are $n_i$ individuals (indexed by $j=1,\ldots,n_i$), for a total of $n=\sum_{i=1}^N n_i$ individuals. We observe data across $K$ intervention stages. 
Let $N_k \le N$ denote the number of clusters randomized at stage $k$. 
At the beginning of each stage $k \in \{1,\dots,K\}$, cluster~$i$ is assigned an intervention $A_{k,i}$. Let $\mathcal{A}_k$ denote the set of feasible intervention options at stage~$k$. 
Let $\bm{S}_{k,i}$ denote a vector of cluster-level covariates that were collected at the end of stage $k$, prior to the assignment of $A_{k+1,i}$.
Baseline (pre-randomization) covariates are denoted by $\bm{S}_{0, i}$.
Let $Y_{i,j}$ denote the individual-level, primary research outcome collected at the end of intervention stage $K$. In this manuscript, we assume $Y_{i,j}$ is a continuous random variable, and we posit that larger values of $Y_{i, j}$ are more desirable. The vector of outcomes in cluster $i$ is denoted by $\bm{Y}_{i} = (Y_{i,1}, \ldots, Y_{i,n_i})$. 
The complete observed data for cluster $i$ can be written in temporal order as:
\begin{equation}
\begin{split}
        \bm{O}_i &=\{ \bm{S}_{0, i}, A_{1,i},\ldots, \bm{S}_{k-1, i}, A_{k,i}, \ldots, \bm{S}_{K-1, i}, A_{K,i}, \bm{Y}_{i}\} \\ &= 
    \{ \bm{H}_{k, i}, A_{k,i}, \ldots, \bm{S}_{K-1, i}, A_{K,i}, \bm{Y}_{i}\}.
\end{split}
\end{equation}
where $\bm{H}_{k,i} = \{\bar{A}_{k-1,i},\,\bar{\bm{S}}_{k-1,i}\}$ denotes the history of observed intervention assignments and candidate tailoring variables measured prior to $A_{k,i}$; hereafter, we use a bar over a variable to denote that variable at the current time and all past values of the same variable. We partition the full history into two parts, $\bm{H}_{k,i} =\{\bm{H}_{k_0,i}, \bm{H}_{k_1,i}\}$, where 
$\bm{H}_{k_1,i}$ contains cluster-level candidate tailoring variables,
and 
$\bm{H}_{k_0,i}$ comprises all other variables that are not correlated with variables in $\bm{H}_{k_1,i}$.
For brevity, we drop the subscript $i$ in $\bm{H}_{k,i}$ or the subscript pair $(i,j)$ in $Y_{i,j}$ when there is no potential for confusion.
Note $\bm{H}_{k}$ comprises all covariates used to define the sequential randomization probabilities. 
For all intervention options $a_k \in \mathcal{A}_k$, 
the randomization probabilities $\pi_k(a_k | \bm{H}_{k}) = \operatorname{Pr}(A_{k}=a_k | \bm{H}_{k})$ are known by design in a cSMART. For simplicity of presentation, we assume that ${A}_k$ is binary (taking values 0 or 1) or singleton given $\bm{H}_{k}$, assigned with probability 1/2 or 0. 

\subsection{Potential Outcomes and Clustered Adaptive Intervention Rules (cAI rules)}

We follow the potential outcome notation in \cite{rubin1974estimating} to define a cAI. Let $\bar{a}_k=(a_1,\ldots,a_k)$ denote a specific history of interventions assigned up to stage~$k$, and $\bm{S}_{k}(\bar{a}_{k})$ denote the vector of cluster-level covariates that had $\bar{a}_{k}$ been assigned in the past up to stage $k$.  
Similarly, let $\bm{Y}(\bar{a}_K)$ denotes the vector of potential individual-level outcomes in a cluster had $\bar{a}_{K}$ been assigned. $\bar{\bm{S}}_{k}(\bar{a}_{k})$ denotes the history of candidate tailoring variables had $\bar{a}_{k}$ been assigned in the past; and note $\bm{H}_{k}(\bar{a}_{k-1}) \equiv (\bar{a}_{k-1},\bar{\bm{S}}_{k-1}(\bar{a}_{k-1}))$. 

A cAI is a sequence of intervention decision rules
$\bar{d}_K = \{d_1,\ldots,d_K\}$, each mapping the history $\bm{H}_{k}(\bar{a}_{k-1})$ to a recommended stage-$k$ treatment $a_k \in \mathcal{A}_k$. 
${Y}_{i,j}(\bar{d}_K)$ is the potential outcome for unit $j$ in cluster $i$ had clustered adaptive intervention $\bar{d}_K\left(\bm{H}_{K}(\bar{a}_{K-1})\right)$ been used to assign treatments $\bar{a}_K$. 
Let $\mathscr{D}_{k}$ be the set of feasible decision rules in stage $k$; and $\bar{\mathscr{D}}_{K} = \cup^K_{k=1} \mathscr{D}_{k}$. 
Define the optimal cAI as $\bar{d}^{opt}_K\left(\bm{H}_{K}(\bar{a}_{k-1})\right) = \argmax_{\bar{d}_{K} \in \bar{\mathscr{D}}_K} E \{ {Y}_{}(\bar{d}_K) \} $.
 


To operationalize the effect of different decision rules on $Y$ and identify the optimal cAI, we adopt the typical 
causal assumptions \cite[][]{robins2004optimal, hernan2010causal} as follows: 

\begin{enumerate}[wide,label=(A\arabic*).,ref=(A\arabic*)]
    \item \textit{Consistency}: For each $k \in \{1,\dots,K\}$, $\bm{S}_{k} \;=\; \bm{S}_{k}(\bar{A}_{k})$ and $\bm{Y} = \bm{Y}(\bar{A}_k)$, where $\bar{A}_{k}$ is the observed sequence of treatments through stage~$k$.
    \item \textit{Sequential Ignorability}: For each and $k \in \{1,\dots,K\}$, the vector of potential covariates $\bigl\{\bm{S}_{\ell}(\bar{a}_{\ell}),\,\ell>k\bigr\}$
    and the potential outcomes $\bm{Y}(\bar{a}_K)$ are independent of $A_{k}$ conditional on treatment history $\bm{H}_{k}$, i.e.,
        $A_{k} \;\perp\!\!\!\perp\; 
   \bigl\{\bm{S}_{k+1}(\bar{a}_{k+1}),\,\dots,\,\bm{S}_{K-1}(\bar{a}_{K-1}),\,\bm{Y}(\bar{a}_K)\bigr\}
   \;\big|\; \bm{H}_{k}$.
   \item \textit{Positivity}: Assume there exist constants $c_0$ and $c_1$ $(0<c_0<c_1<1)$, such that the propensity score $\pi_k(a_k|\bm{H}_{k}) = \operatorname{Pr}(A_{K} = a_K|\bmath{H}_{K}) \in (c_0, c_1)$ with probability 1.
\end{enumerate}
Note that in a cSMART, the sequential randomization probabilities $\pi_k(a_k|\bm{H}_{k})$ are known; thus, Assumptions A2 and A3 are ensured by design. 

\section{Clustered Q-learning Framework}
\subsection{Q-functions and Optimal cAI}

To build intuition for the method of estimating an optimal cAI, we adopt the \emph{$Q$-learning} technique \cite[][]{clifton2020q}, which uses ideas from dynamic programming. 
Conceptually, $Q$-learning proceeds backward from the final stage to the first. We define the stage $K$ $Q$-function as:
\vspace{-3pt}
\begin{equation}
Q_K(\bm{H}_{K},a_K) = \mathbb{E} \{{Y}({a}_K) | \bm{H}_{K}\} = \mathbb{E} \{ {Y} | \bm{H}_{K},{A}_K={a}_K\},
\end{equation}
\vspace{-35pt}

which is the expected final outcome if we assign treatment $a_K$ at stage~$K$ after having observed the past treatment and covariate history $\bm{H}_{K}$. Note the second equality is based on assumptions A1 and A2. 
Then, the optimal decision rule $d_K^{opt}$ chooses the $a_K$ maximizing the stage $K$ $Q$-function:
\vspace{-3pt}
\begin{equation}
 d_{K}^{opt}(\bm{H}_{K}) = \argmax_{\,a_{K} \,\in\,\mathcal{A}_{K}\!} \;
   Q_{K}\bigl(\bm{H}_{K},\,a_K\bigr).
\end{equation}
\vspace{-33pt}

Then, we define $\widetilde{Y}_{K-1} \mathrel{\overset{\triangle}{=}} \max_{\,a_{K} \,\in\,\mathcal{A}_{K}\!} Q_{K}\left(\bm{H}_{K},a_K\right)$ as a \textit{pseudo-outcome} in the previous stage $K-1$; that is, the potential outcome given the last stage following the treatment assigned by the optimal decision rule $d_K^{opt}$.
Generally, for each earlier stage $k<K$, $Q$-learning uses a outcome recursively updated that assumes future treatments to be optimized. 
Concretely, we define:
\vspace{-3pt}
\begin{equation}
Q_k(\bm{H}_{k},a_k) = \mathbb{E} \{ \widetilde{Y}_{k} | \bm{H}_{k},{A}_k={a}_k\},
\end{equation}
\vspace{-33pt}

where $\widetilde{Y}_k \mathrel{\overset{\triangle}{=}} \max_{\,a_{k+1} \,\in\,\mathcal{A}_{k+1}\!} Q_{k+1}\left(\bm{H}_{k+1},a_{k+1}\right)$ is the stage-$k$ \textit{pseudo-outcome}, which is the potential outcome given all future decisions from stage $k+1$ onward are chosen optimally. These $\widetilde{Y}_1 \ldots, \widetilde{Y}_{K-1}$ are also named as ``optimal benefit-to-go functions" in \cite{murphy2003optimal}. 
Then, we have:
\vspace{-5pt}
\begin{equation}
 d_{k}^{opt}\bigl(\bm{H}_{k}\bigr)
 \;=\;
 \argmax_{\,a_{k} \,\in\,\mathcal{A}_{k}\!}
 \;
 Q_{k}\bigl(\bm{H}_{k},\,a_k\bigr).
\end{equation}
\vspace{-30pt}

With the $Q$-function for each $k \in \{1, \ldots, K\}$, the overall optimal cAI $\bar{d}_K^{opt}=\bigl(d_1^{opt},\,\dots,\,d_K^{opt}\bigr)$ can be estimated as in the following section.

\subsection{Estimation}
\label{S:estimation}

To model the $Q$-functions for all stages, we utilize generalized regression on $H_{k_0}, H_{k_1}, A_k H_{k_1}$, and estimate regression coefficients via
estimation equations 
for ease of implementation and interpretability. Further, it enables us to address, in this first paper on the topic, challenges related to developing methods for making statistical inferences.
Specifically, we model the stage $k$ $(k \in \{1, \ldots, K\})$ $Q$-function by:
\vspace{-3pt}
\begin{equation}
\begin{split}
        Q_{K}\left(\bm{H}_{k}, A_{k}; {\bm{\gamma}}_k, \bm{\beta}_k, \Psi_k\right) &= {\bm{\gamma}}_k^\intercal \bm{H}_{k_0} + {\bm{\beta}}_k^\intercal \bm{H}_{k_1} + (\bm{\Psi}_k^\intercal \bm{H}_{k_1}) A_{k},
\end{split}
\end{equation}
\vspace{-30pt}

where $\bm{\beta}_k$ denotes the coefficients for the cluster-level candidate tailoring variables in $\bm{H}_{k_1}$, $\bm{\gamma}_k$ denotes the coefficients for the other variables in $\bm{H}_{k_0}$, and $\bm{\Psi}_k$ denotes the causal effect of treatment conditional on variables in $\bm{H}_{k_1}$. Let $\bm\theta_{k} = \left({\bm{\gamma}}_k^\intercal, \bm{\beta}_k^\intercal, \bm{\Psi}_k^\intercal \right)^\intercal$ denote the vector of all coefficients.

In standard cluster-randomized trials, analysts commonly employ working models that decompose the total residual variance into cluster- and individual-level components.
Such working models account for the non-independence of observations within each cluster, often via an intra-cluster correlation coefficient (ICC) \cite[][]{campbell2004intracluster}, and can confer important interpretational and efficiency benefits.
Here, we generalize these ideas to linear $Q$-learning. 
Fix a stage $k$. Suppose we have a pseudo-outcome $\widetilde{Y}_{k,i,j}$ for individual $j$ in cluster $i$ and stage $k$ model $Q_k(\cdot;\theta_k)$. Define the residual as  $\varepsilon_{k,i,j}(\bm{H}_{k},\,A_k;\,\bm\theta_k) = \bigl(\,\widetilde{Y}_{k,i,j}(A_k) -Q_k(\bm{H}_{k},\,A_k;\bm\theta_k)\bigr)$ and assume $\varepsilon_{k, i, j} \perp \varepsilon_{k, i^{\prime}, j}, \forall j, i \neq i^{\prime}$.
We can posit the working decomposition:
\vspace{-3pt}
\begin{equation}
  \varepsilon_{k,i,j} = \alpha_{k,i} \;+\;\nu_{k,i,j},
  \quad
  \sigma^2_{k,\alpha} = \mathrm{Var}(\alpha_{k,i}), 
  \quad
  \sigma^2_{k,\nu}=\mathrm{Var}(\nu_{k,i,j}).
\end{equation}
\vspace{-30pt}

Then the total residual variance is $\sigma^2_{k,\varepsilon} = \sigma^2_{k,\alpha} + \sigma^2_{k,\nu}$, and the ICC is $\rho_k = \sigma^2_{k,\alpha}/\sigma^2_{k,\varepsilon}$. 
Thus, one might choose $\mathrm{Var}(\widetilde{\bm{Y}}_{k,i}) \approx \sigma^2_{k,\varepsilon} \,\bm{J}_k$,
where $\bm{J}_k$ is an $n_i\times n_i$ matrix with 1 on the diagonal and $\rho_k$ in the off-diagonals. 
The within-cluster error for cluster $i$ can be expressed as $\bm\varepsilon_{k, i}=\left(\varepsilon_{k, i, 1}, \varepsilon_{k, i, 2}, \ldots, \varepsilon_{k, i, n_i}\right)^\intercal$. 
We assume $E\left({\bm\varepsilon}_{k, i} | \bm{H}_{k, i j}\right)={\bm{0}}$ and let $\boldsymbol{\Sigma}_{k,i}$
be a model for the $n_i \times n_i$ intra-cluster variance-covariance matrix $\operatorname{Cov}(\bm\varepsilon_{k,i} |  \bm{H}_{k}$). 

Our scientific interest lies in evaluating whether candidate cluster-level tailoring variables are useful in defining a cluster adaptive intervention. Thus, we will only focus on inferences about $\bm\Psi_k$.
The term $\bm\Psi_{k}^\intercal  \bm{H}_{k_1} $, in particular, potentially defines the form of the estimated cAI.
The estimation procedure involves $K$ regressions conducted in reverse order, i.e., beginning with stage $K$ and ending with stage 1.
\begin{enumerate}[leftmargin=20pt, rightmargin=0pt,itemindent=0em, itemsep = 0em, parsep = 0em, labelwidth = 5em]
    \item For stage $K$, 
    estimate parameters $ \theta_{K} = \left({\bm{\gamma}}_K^\intercal, \bm{\beta}_K^\intercal, \bm{\Psi}_K^\intercal \right)^\intercal$ by minimizing residual sum of squares between the outcome $Y_{i j}$ and the $Q$-functions $Q_{K, i, j}(\bm{H}_{K, i, j}, A_{K, i})$ as:  
    \begin{equation}
        \widehat {\bm{\theta}}_K = \argmin_{\bm{\theta}_K} 
        \sum_{i=1}^{N_K}
    \left[\bm{Y}_{i}- \bm{Q}_{K,i}\right]^\intercal \boldsymbol{\Sigma}_{K, i}^{-1}
    \left[\bm{Y}_{i}- \bm{Q}_{K,i}\right],
    \end{equation}
    where $ \bm{Q}_{K,i} = \big(Q_{K, i, 1}(\bm{H}_{K, i, 1}, A_{K, i}),
    \ldots, Q_{K, i, n_i}(\bm{H}_{K, i, n_i}, A_{K, i}) \big)^\intercal,$ 
    and $\bm{Y}_i=\left(Y_{i 1}, \ldots, Y_{i n_i}\right)^\intercal.$ 
    When $\operatorname{Cov}\left({\bm\varepsilon}_{K, i} | \bm{H}_{K, i}\right)$ is unknown, we model it with a working variance-covariance matrix 
    $\boldsymbol{\Sigma}_{K,i}$.
    \\ \vspace{-20pt}
    \item For stages $k = \{K-1, \ldots, 1\}$, perform the following steps backward from stage $K-1$ to stage 1:
    \begin{enumerate}[label=(\roman*), leftmargin=17pt, 
    rightmargin=0pt,
    itemindent=0em, 
    itemsep = 0em, 
    parsep = 0em, 
    labelwidth = 5em]
        \item Calculate the stage $k$ specific pseudo-outcome $\widetilde{Y}_{k, i, j}$. Noting that the pseudo-outcome can be defined as a function of the final observed outcome $Y_{i j}$ and intermediate observed outcomes $Y_{1, i, j}, \ldots, Y_{K - 1, i, j}$. Since we assume a scenario that solely observes the final stage outcome, the stage $k$ pseudo-outcome can then be defined as:
\begin{equation}
\widetilde{Y}_{k, i, j} =
\sup _{d_{k,i} \in \mathscr{D}_{k}} \left\{ \widehat{\bm{\gamma}}_{k+1}^\intercal \bm{H}_{{k+1}_0, i, j} + \widehat{\bm{\beta}}_{k+1}^\intercal \bm{H}_{{k+1}_1, i} + ( \widehat{\bm{\Psi}}_{k+1}^\intercal \bm{H}_{{k+1}_1, i} ) d_{k+1, i} \right\}.
\end{equation}
\item Estimate parameters $ \theta_{k} = \left({\bm{\gamma}}_k^\intercal, \bm{\beta}_k^\intercal, \bm{\Psi}_k^\intercal \right)^\intercal$ by minimizing residual sum of squares between the pseudo-outcome $\widetilde Y_{k, i j}$ and the $Q$-functions $Q_{k, i j}(\bm{H}_{k, i j}, A_{k, i})$:  \begin{equation}
        \widehat {\bm{\theta}}_k = \argmin_{\bm{\theta}_k} 
        \sum_{i=1}^{N_k}
    \left[ \widetilde{\bm{Y}}_{k,i} - \bm{Q}_{k,i} \right]^\intercal 
    \boldsymbol{\Sigma}_{k, i}^{-1} 
    \left[ \widetilde{\bm{Y}}_{k,i} - \bm{Q}_{k,i} \right],
    \end{equation}
    where $ \bm{Q}_{k,i} = (Q_{k, i, 1}(\bm{H}_{k, i, 1}, A_{k, i}),
    \ldots, Q_{k, i, n_i} (\bm{H}_{k, i n_1}, A_{k, i}))^\intercal,$ and \\ $\widetilde{\bm{Y}}_{k, i}=\left(\widetilde Y_{k,i, 1}, \ldots, \widetilde Y_{k,i, n_i}\right)^\intercal$, and
   $\boldsymbol{\Sigma}_{k, i}$ is a working variance-covariance matrix.
\end{enumerate}
\end{enumerate}

\subsection{Constructing Confidence Intervals}

Conducting inference in a single-stage estimation with correlated data is straightforward as standard regression techniques that account for the correlation of outcomes within clusters are used for estimation and inference. Estimation in a multi-stage setting, however, can be problematic. Specifically, because estimation proceeds in a backward manner and, the $Q$-learning approach involves regression on a pseudo-outcome that accounts for the optimal intervention received at all future stages, estimation at all stages prior to the last may involve maximization of a non-smooth function, known as ``non-regularity". 
Refer to Web Appendix B for an expanded discussion of non-regularity and an illustration of non-regularity in the context of a two-stage, two-treatment setting.

The problem of non-regularity is well-described in the literature \cite[][]{robins2004optimal} and multiple solutions have been developed to account for non-regular estimators, including threshold estimators \cite[][]{chakraborty2010inference, moodie2010estimating} and adaptive confidence intervals \cite[][]{laber2014dynamic}. Another approach for estimating confidence intervals of non-regular parameters used with standard $Q$-learning is the $m$-out-of-$n$ bootstrap \cite[][]{chakraborty2013inference}.

We propose a $M$-out-of-$N$ Cluster Bootstrap for clustered data, unifying the Cluster Bootstrap and the $m$-out-of-$n$ standard bootstrap 
for estimating confidence intervals for parameters indexing the optimal cAI in the setting of clustered $Q$-learning with an assumed parametric, linear structure. Notably, our method relies on a bootstrap method for estimating confidence intervals, in contrast to methods that utilize traditional, robust regression-based standard errors, which have demonstrated divergences from nominal confidence interval coverage under conditions of non-regularity. As is generally well-known, the standard Cluster Bootstrap performs resampling at the cluster level rather than the individual level, which is critical when estimating the degree of variability of an estimator in the presence of correlated data \cite[][]{Hox_2010}. It has been shown that if assuming the number of clusters is large \cite[][]{field2007bootstrapping}, model errors are uncorrelated across clusters but correlated within clusters, clusters are exchangeable \cite[][]{bouwmeester2013internal}, and the empirical distribution $F_N(x)$ is a reasonable approximation to the underlying distribution $F(x)$, the Cluster Bootstrap is then asymptotically consistent \cite[][]{cheng2013cluster}. 

To address the issue of non-regularity, 
we select a resample size of $M$ from $N$ clusters, with $M \leqslant N$, which reflects the degree of non-regularity in the underlying generative model. 
In specific, for stage $k \in \{1, \ldots, K-1\}$, 
the degree of non-regularity ($p_k$) corresponds to 
the stage $k+1$ intervention effect of the prescriptive cluster-level covariates $\bm{H}_{k+1_1}$ is defined as 
\vspace{-5pt}
\begin{equation}
    p_k \overset{\Delta}{=} \operatorname{Pr} \{\bm{H}_{k+1_1}: { \bm\Psi}_{k+1}^\intercal \bm{H}_{k+1_1} =0\}.
\end{equation}
\vspace{-30pt}

Extended from \cite{chakraborty2013inference}, to estimate the resample size at stage $k$, denoted as $M_k$, a T-statistic is calculated for each of $N_{k+1}$ clusters as $T_{k, i}=\frac{\widehat{\bm{\Psi}}_{k+1}^\intercal \bm{H}_{k+1_1, i} }{\widehat{\operatorname{SE}}\left(\widehat{\bm{\Psi}}_{k+1}^\intercal \bm{H}_{k+1_1, i} \right)}$, with the standard errors derived using the sandwich variance estimator. Utilizing a pre-defined threshold $\eta$, we estimate $p_k$ as the proportion of clusters with a T-statistic below $\eta$: 
\vspace{-0pt}
\begin{equation}
\widehat{p}_{k}=\frac{1}{N_{k+1}} \sum_{i=1}^{N_{k+1}} \operatorname{I} \left(| T_{k, i}| \leqslant \eta\right).
\end{equation}
\vspace{-30pt}

To restrict the selection of $M_k$ into a smaller class, we define $\widehat{M}_k \overset{\Delta}{=} N_k^{f\left(\widehat{p}_{k}\right)}$, where $f(\cdot)\in (0,1]$ is a continuous and monotone decreasing function with bounded first-order derivative, and we assume $f(0)=1$. Correspondingly, we propose to use a function $f\left(p_k\right)=\frac{1+\lambda\left(1-p_k\right)}{1+\lambda}$ with a tuning parameter $\lambda$. Then, the resample size $M_k$ at stage $k$ is formulated as
\vspace{-0pt}
\begin{equation}
    \widehat{M}_k= N_k^{\frac{1+\lambda-\lambda \widehat{p}_{k}}{1+\lambda}}.
\end{equation}
\vspace{-38pt}

In a highly regular example, i.e., when there is a strong stage $k+1$ intervention effect for all clusters, the degree of non-regularity is $p_{k} = 0$; consequently, $M_k = N_k$, which represents the standard Cluster Bootstrap. With increasingly higher degrees of non-regularity for the stage $k$ estimation, the value of $M_k$ decreases relative to $N_k$. With a desired global Type-I error rate $\alpha$ and considering a Bonferroni correction for multiple hypothesis tests, a natural choice for the threshold $\eta$ is $t_{n_i-1,1-\frac{\alpha}{2N_{k+1}}}$.
Empirically, value of tuning parameter $\lambda$ can be chosen in a range of $0.025 - 0.10$, or selected via a data-driven algorithm (see Web Appendix A). In addition, one could explore more flexible data-driven strategies for selecting the tuning parameter that further balance coverage and interval length.


\subsection{Implementation}
To facilitate the application of the proposed method by users, we provide a comprehensive tutorial that exemplifies the analysis of a two-stage cSMART in Web Appendix A. This tutorial meticulously outlines the process of selecting variables and tuning parameters, constructing $Q$-functions, practical considerations on variance estimation, and making inferences about the tailoring effect at the first stage with the $M$-out-of-$N$ Cluster Bootstrap technique.
The R code for implementing the algorithm is accessible with the Web Appendix Materials, and an R package \texttt{clusterQ} is available at {https://github.com/SelinaSong0412/clusterQ}.

\section{Simulation Studies}

\subsection{Aim 1: Performance Across Different Sample Size and Effect Size Settings}

To evaluate whether the proposed method can estimate parameters corresponding to multi-stage candidate tailoring variables with a low degree of bias and near-nominal coverage under varying degrees of (non-)regularity, we generate data for a two-stage, unrestricted cSMART (i.e., type I design).
We consider three sample size scenarios:
(Scenario~1) $N = 80$ clusters, each with $n_i = 20$ individuals (large number of clusters); 
(Scenario~2) $N = 20$ clusters, each with $n_i = 80$ individuals (small number of clusters); and
(Scenario~3) $N = 30$ clusters with varying numbers of individuals per cluster $n_i \sim \mathrm{Unif}(10,30)$. This setup reflects real-world cSMART designs, which often have moderate cluster counts and heterogeneous cluster sizes 
\cite[][]{kilbourne2014protocol, kilbourne2018adaptive, quanbeck2020balanced, pan2024finite}.

For all scenarios, one binary candidate cluster-level tailoring variable $X_1 \in \{1, -1\}$ is generated at baseline. Randomization to the first-stage cluster-level binary intervention $A_1 \in \{1, -1\}$ occurs in a ratio of 1:1. 
An intermediate binary response $X_2$ (used as a second-stage candidate tailoring variable) depends on $(X_1, A_1)$.
The second-stage intervention $A_2 \in \{1, -1\}$ is again assigned with equal probability. 
The final individual-level outcome observed following the second stage, denoted as $Y$, is continuous and approximately normally distributed, with larger values preferable. 
Within-cluster correlation is imposed using an ICC coefficient $\rho \in \{0.05, 0.1, 0.2\}$ to capture different degrees of within-cluster dependence. 
This data-generating scheme allows the true stage~1 parameters, $\Psi_1$, to exist in closed form (derivation provided in Web Appendix C), so that bias and coverage are straightforwardly computed.

Following \cite{chakraborty2010inference} and \cite{laber2014dynamic}, we design six distinct data-generating examples (Ex.~1--6) that reflect regular or non-regular conditions and three different underlying first-stage effect sizes (refer to Table~Web 1 in Web Appendix). The former is to test the proposed method's performance in dealing with non-regularity as well as its stability in regular cases; the latter is to examine the method's ability to handle different real-world cSMART scenarios.
Specifically, Ex.~1--3 represent regular conditions ($p=0$) with first-stage effect sizes 0.2, 0.5, and 0.8, reflecting small, moderate, and large effect sizes, respectively. Ex.~4--6 represent non-regular conditions ($p \neq 0$) again with first-stage effect sizes 0.2, 0.5, and 0.8. (See Web Appendix C.2 for full details on the data generation, how non-regularity arises, and the underlying clinical assumptions).

For each Scenario--Ex. pair, we compare the performance of: 
(i) the proposed $M$-out-of-$N$ Cluster Bootstrap (MN-CB), 
(ii) the standard Cluster Bootstrap (CB; i.e., full resampling from all $N$ clusters),
and 
(iii) the $m$-out-of-$n$ Bootstrap proposed by \cite{chakraborty2013inference} (mn-B). 
The mn-CB deals with the non-regularity issue but assumes independent outcomes; the standard CB deals with the clustering effect but does not handle the non-regularity; while the proposed MN-CB is designed to deal with both problems.
In addition, for all simulations related to Scenario 3, we add a fourth method for comparison -- (iv) the proposed $M$-out-of-$N$ Cluster Bootstrap with wrong (independent) working correlation model (MN-CB-w). This additional comparison is only added to Scenario 3 because it has been shown in the previous literature that the working correlation may affect the inference only when the number of patients in each group is not fixed \cite[][]{pan2024finite}, which is also proven by our simulation. 
Further, the comparison of the estimation efficiency of the MN-CB with different working correlation models is discussed in Aim 2 in Section \ref{simulation:aim2}.
All methods use fixed parameter values of $\lambda = 0.025$ and $\eta = t_{n_i-1, 1-0.001}$.
Performance metrics include estimation bias, stand error (SE), and coverage rate and the length of the 95\% confidence interval of $\widehat \Psi_1$, each is estimated over $B = 500$ Monte Carlo iterations with 1000 bootstrap samples per iteration.



\textbf{Scenario 1 ($\bm{N = 80}$, $\bm{n_i = 20}$).}
Across Ex.~1--6 (Table~\ref{tab:tab1}), the proposed MN-CB achieves near-nominal or slightly conservative coverage in most cells, consistently outperforming both CB and mn-B in the large majority of settings. For instance, in the regular examples, MN-CB coverage remains above 95\%, even for higher ICC $\rho = 0.2$. By contrast, mn-B tends to show a modest drop in coverage as $\rho$ increases from 0.05 to 0.2 or as the stage~1 effect size grows from 0.2 to 0.8. The standard CB method performs reasonably well under regular conditions, often achieving coverage close to MN-CB. In the non-regular examples, MN-CB has a clearer advantage over both comparators; CB and mn-B exhibit notable undercoverage in certain settings, whereas MN-CB remains stable at or near the nominal level. Overall, these results indicate that MN-CB yields excellent coverage in larger-sample cSMARTs, handling both regular and non-regular conditions reliably.            

\begin{table}
\centering
\resizebox{\columnwidth}{!}{%
\begin{tabular}{ccclcccccccc}
\hline \hline
\multirow{2}{*}{\textbf{\begin{tabular}[c]{@{}c@{}}Regularity \\ Setting\end{tabular}}} & \multirow{2}{*}{\textbf{Ex.\#}} & \multirow{2}{*}{\textbf{\begin{tabular}[c]{@{}c@{}}Stage 1\\ Effect Size\end{tabular}}} & \multirow{2}{*}{} & \multicolumn{2}{c}{$\bm\rho$\textbf{ = 0.05}}       & \textbf{}            & \multicolumn{2}{c}{$\bm\rho$\textbf{ = 0.1}}        & \textbf{}            & \multicolumn{2}{c}{$\bm\rho$\textbf{ = 0.2}}        \\ \cline{5-6} \cline{8-9} \cline{11-12} 
                                 &                                                                                              &                                                                                         &                   & Coverage (\%)        & Length               &                      & Coverage (\%)        & Length               &                      & Coverage (\%)        & Length         \\  \hline \\
\multirow{11}{*}{\begin{tabular}[c]{@{}c@{}}Regular\\ ($p = 0$)\end{tabular}} & \multirow{3}{*}{1} & \multirow{3}{*}{0.2}                                                                    & mn-B              & 93.0                 & 0.121 (0.007)        &                      & 86.2                 & 0.120 (0.007)        &                      & 72.8                 & 0.118 (0.008)        \\
                                 &                                                                                              &                                                                                         & CB                & 96.4                 & 0.144 (0.015)        &                      & 96.4                 & 0.175 (0.019)        &                      & 96.6                 & 0.225 (0.024)        \\
                                 &                                                                                              &                                                                                         & MN-CB             & \textbf{96.6}        & 0.147 (0.016)        &                      & \textbf{96.6}        & 0.180 (0.020)        &                      & \textbf{96.8}        & 0.231 (0.025)        \\
                                 & \multicolumn{1}{l}{}                                                                         & \multicolumn{1}{l}{}                                                                    &                   &                      &                      &                      &                      &                      &                      &                      &                      \\
                                 & \multirow{3}{*}{2}                                                                           & \multirow{3}{*}{0.5}                                                                    & mn-B              & 90.6                 & 0.120 (0.007)        &                      & 77.4                 & 0.119 (0.007)        &                      & 66.4                 & 0.117 (0.008)        \\
                                 &                                                                                              &                                                                                         & CB                & 95.4                 & 0.143 (0.015)        &                      & 95.4                 & 0.175 (0.019)        &                      & 95.4                 & 0.225 (0.024)        \\
                                 &                                                                                              &                                                                                         & MN-CB             & \textbf{95.8}        & 0.147 (0.016)        &                      & \textbf{96.2}        & 0.180 (0.020)        &                      & \textbf{96.4}        & 0.230 (0.025)        \\
                                 & \multicolumn{1}{l}{}                                                                         & \multicolumn{1}{l}{}                                                                    &                   &                      &                      &                      &                      &                      &                      &                      &                      \\
                                 & \multirow{3}{*}{3}                                                                           & \multirow{3}{*}{0.8}                                                                    & mn-B              & 90.6                 & 0.121 (0.007)        &                      & 84.4                 & 0.120 (0.007)        &                      & 71.0                 & 0.117 (0.008)        \\
                                 &                                                                                              &                                                                                         & CB                & 95.8                 & 0.144 (0.015)        &                      & 95.8                 & 0.175 (0.019)        &                      & 95.8                 & 0.226 (0.024)        \\
                                 &                                                                                              &                                                                                         & MN-CB             & \textbf{96.2}        & 0.148 (0.016)        &                      & \textbf{96.0}        & 0.180 (0.020)        &                      & \textbf{96.0}        & 0.232 (0.025)       \bstrut \\  \hline 
\multicolumn{1}{l}{}             & \multicolumn{1}{l}{}                                                                         & \multicolumn{1}{l}{}                                                                    &                   & \multicolumn{1}{l}{} & \multicolumn{1}{l}{} & \multicolumn{1}{l}{} & \multicolumn{1}{l}{} & \multicolumn{1}{l}{} & \multicolumn{1}{l}{} & \multicolumn{1}{l}{} & \multicolumn{1}{l}{} \\
\multirow{11}{*}{\begin{tabular}[c]{@{}c@{}}Non-regular\\ ($p \neq 0$)\end{tabular}} & \multirow{3}{*}{4} & \multirow{3}{*}{0.2}                                                                    & mn-B              & 94.4                 & 0.143 (0.008)        &                      & 87.0                 & 0.141 (0.009)        &                      & 74.2                 & 0.138 (0.011)        \\
                                 &                                                                                              &                                                                                         & CB                & 94.4                 & 0.145 (0.016)        &                      & 94.4                 & 0.177 (0.019)        &                      & 94.4                 & 0.227 (0.024)        \\
                                 &                                                                                              &                                                                                         & MN-CB             & \textbf{96.0}        & 0.153 (0.017)        &                      & \textbf{95.8}        & 0.187 (0.021)        &                      & \textbf{95.4}        & 0.241 (0.027)        \\
                                 & \multicolumn{1}{l}{}                                                                         & \multicolumn{1}{l}{}                                                                    &                   &                      &                      &                      &                      &                      &                      &                      &                      \\
                                 & \multirow{3}{*}{5}                                                                           & \multirow{3}{*}{0.5}                                                                    & mn-B              & 90.8                 & 0.122 (0.007)        &                      & 80.8                 & 0.121 (0.008)        &                      & 66.0                 & 0.119 (0.008)        \\
                                 &                                                                                              &                                                                                         & CB                & 95.8                 & 0.146 (0.015)        &                      & 95.8                 & 0.178 (0.019)        &                      & 95.8                 & 0.229 (0.024)        \\
                                 &                                                                                              &                                                                                         & MN-CB             & \textbf{96.4}        & 0.149 (0.016)        &                      & \textbf{96.0}        & 0.182 (0.019)        &                      & \textbf{96.0}        & 0.235 (0.025)        \\
                                 & \multicolumn{1}{l}{}                                                                         & \multicolumn{1}{l}{}                                                                    &                   &                      &                      &                      &                      &                      &                      &                      &                      \\
                                 & \multirow{3}{*}{6}                                                                           & \multirow{3}{*}{0.8}                                                                    & mn-B              & 89.2                 & 0.122 (0.007)        &                      & 80.6                 & 0.121 (0.007)        &                      & 68.4                 & 0.119 (0.008)        \\
                                 &                                                                                              &                                                                                         & CB                & 94.4                 & 0.145 (0.016)        &                      & \textbf{94.2}        & 0.177 (0.019)        &                      & \textbf{94.4}        & 0.227 (0.024)        \\
                                 &                                                                                              &                                                                                         & MN-CB             & \textbf{94.6}        & 0.149 (0.017)        &                      & \textbf{94.2}        & 0.182 (0.020)        &                      & \textbf{94.4}        & 0.234 (0.025)        \bstrut \\  \hline \hline
\end{tabular}
}
\caption{Results for simulation Scenario 1 (Large number of clusters). Estimates of 95\% confidence interval coverage and length for the coefficient of the $X_1 A_1$ interaction effect, $\psi_{11}$, estimated in the first-stage estimation for $N$ = 80 clusters with $n_i$ = 20 individuals per cluster. Largest coverage for each setting is shown in bold font. $\rho$ refers to the intra-cluster correlation (ICC) used to generate the simulation data; $p$ refers to the degree of non regularity; Coverage (\%) represents the coverage of the 95\% confidence interval; Length represents the length of the 95\% confidence interval. Abbreviations: mn-B: $m$-out-of-$n$ Bootstrap; CB: Cluster Bootstrap; MN-CB: the proposed $M$-out-of-$N$ Cluster Bootstrap.}
\label{tab:tab1}
\end{table}

\textbf{Scenario 2 ($\bm{N = 20}$, $\bm{n_i = 80}$).}
We observe slightly higher bias (still negligible, with a maximum around 0.01) and standard error of estimation across all examples; both increase as ICC increases (see Web Appendix C). 
As shown in Table~\ref{tab:tab2}, coverage decreases somewhat for all methods compare to that in Scenario 1, as fewer clusters reduce effective sample information.
Notably, for the regular Ex.~1--3, CB sometimes matches or slightly exceeds MN-CB coverage for smaller or moderate stage~1 effect sizes. This is most pronounced under lower ICC of $\rho = 0.05$ or $\rho = 0.1$, where the difference can be a few percentage points in favor of CB. However, as the first-stage effect size grows large or in non-regular examples, MN-CB again generally provides more robust coverage, often hitting or staying near the 95\% target. This is not surprising, since there may be a cost to resampling fewer clusters under regular scenarios relative to full resampling (CB). An interesting observation from our simulations is that this ``no free lunch” phenomenon becomes less evident in non-regular settings with more clusters and larger ICC.


\begin{table}
\centering
\resizebox{\columnwidth}{!}{%
\begin{tabular}{ccclcccccccc}
\hline  \hline
 \multirow{2}{*}{\textbf{\begin{tabular}[c]{@{}c@{}}Regularity \\ Setting\end{tabular}}}   & \multirow{2}{*}{\textbf{Ex. \#}}   & \multirow{2}{*}{\textbf{\begin{tabular}[c]{@{}c@{}}Stage 1\\ Effect Size\end{tabular}}} & \multirow{2}{*}{} & \multicolumn{2}{c}{$\bm\rho$\textbf{ = 0.05}}       & \textbf{}            & \multicolumn{2}{c}{$\bm\rho$\textbf{ = 0.1}}        & \textbf{}            & \multicolumn{2}{c}{$\bm\rho$\textbf{ = 0.2}}        \\ \cline{5-6} \cline{8-9} \cline{11-12} 
                                 &                                                                                              &                                                                                         &                   & Coverage (\%)        & Length               &                      & Coverage (\%)        & Length               &                      & Coverage (\%)        & Length                \\  \hline \\
\multirow{11}{*}{\begin{tabular}[c]{@{}c@{}}Regular\\ ($p = 0$)\end{tabular}}    & \multirow{3}{*}{1}                           & \multirow{3}{*}{0.2}                                                                    & mn-B              & 69.8                 & 0.141 (0.022)        &                      & 55.3                 & 0.136 (0.021)        &                      & 42.4                 & 0.130 (0.020)        \\
                                 &                                                                                              &                                                                                         & CB                & \textbf{92.9}        & 0.289 (0.083)        &                      & \textbf{92.9}        & 0.387 (0.110)        &                      & 92.9                 & 0.532 (0.153)        \\
                                 &                                                                                              &                                                                                         & MN-CB             & 91.6                 & 0.288 (0.082)        &                      & 92.4                 & 0.388 (0.112)        &                      & \textbf{93.1}        & 0.535 (0.154)        \\
                                 &                                               \multicolumn{1}{l}{}                           & \multicolumn{1}{l}{}                                                                    &                   &                      &                      &                      &                      &                      &                      &                      &                      \\
                                 &                                               \multirow{3}{*}{2}                             & \multirow{3}{*}{0.5}                                                                    & mn-B              & 69.2                 & 0.142 (0.024)        &                      & 53.4                 & 0.137 (0.023)        &                      & 39.3                 & 0.131 (0.022)        \\
                                 &                                                                                              &                                                                                         & CB                & \textbf{92.3}        & 0.285 (0.081)        &                      & \textbf{92.2}        & 0.383 (0.110)        &                      & 92.1                 & 0.525 (0.151)        \\
                                 &                                                                                              &                                                                                         & MN-CB             & 92.1                 & 0.286 (0.084)        &                      & 91.8                 & 0.382 (0.109)        &                      & \textbf{92.2}        & 0.526 (0.148)        \\
                                 &                                               \multicolumn{1}{l}{}                           & \multicolumn{1}{l}{}                                                                    &                   &                      &                      &                      &                      &                      &                      &                      &                      \\
                                 &                                               \multirow{3}{*}{3}                             & \multirow{3}{*}{0.8}                                                                    & mn-B              & 64.7                 & 0.142 (0.022)        &                      & 48.8                 & 0.136 (0.022)        &                      & 37.1                 & 0.130 (0.021)        \\
                                 &                                                                                              &                                                                                         & CB                & 92.1                 & 0.278 (0.074)        &                      & 92.1                 & 0.372 (0.100)        &                      & \textbf{91.9}        & 0.511 (0.137)        \\
                                 &                                                                                              &                                                                                         & MN-CB             & \textbf{92.3}        & 0.278 (0.075)        &                      & \textbf{92.5}        & 0.373 (0.100)        &                      & \textbf{91.9}        & 0.512 (0.138)         \bstrut \\  \hline
\multicolumn{1}{l}{}             & \multicolumn{1}{l}{}                                                                         & \multicolumn{1}{l}{}                                                                    &                   & \multicolumn{1}{l}{} & \multicolumn{1}{l}{} & \multicolumn{1}{l}{} & \multicolumn{1}{l}{} & \multicolumn{1}{l}{} & \multicolumn{1}{l}{} & \multicolumn{1}{l}{} & \multicolumn{1}{l}{} \\

\multirow{11}{*}{\begin{tabular}[c]{@{}c@{}}Non-regular\\ ($p \neq 0$)\end{tabular}}   & \multirow{3}{*}{4}                     & \multirow{3}{*}{0.2}                                                                    & mn-B              & 72.7                 & 0.165 (0.030)        &                      & 56.6                 & 0.155 (0.03)         &                      & 38.6                 & 0.143 (0.028)        \\
                                 &                                                                                              &                                                                                         & CB                & 90.2                 & 0.279 (0.083)        &                      & 90.2                 & 0.375 (0.111)        &                      & 89.7                 & 0.514 (0.152)        \\
                                 &                                                                                              &                                                                                         & MN-CB             & \textbf{91.1}        & 0.288 (0.086)        &                      & \textbf{91.1}        & 0.386 (0.117)        &                      & \textbf{90.9}        & 0.533 (0.159)        \\
                                 &                                             \multicolumn{1}{l}{}                             & \multicolumn{1}{l}{}                                                                    &                   &                      &                      &                      &                      &                      &                      &                      &                      \\
                                 &                                              \multirow{3}{*}{5}                              & \multirow{3}{*}{0.5}                                                                    & mn-B              & 69.0                 & 0.144 (0.024)        &                      & 52.9                 & 0.140 (0.024)        &                      & 40.9                 & 0.133 (0.023)        \\
                                 &                                                                                              &                                                                                         & CB                & 91.7                 & 0.285 (0.088)        &                      & 92.2                 & 0.383 (0.119)        &                      & 91.3                 & 0.524 (0.160)        \\
                                 &                                                                                              &                                                                                         & MN-CB             & \textbf{92.8}        & 0.285 (0.089)        &                      & \textbf{92.4}        & 0.382 (0.117)        &                      & \textbf{91.6}        & 0.525 (0.162)        \\
                                 &                                             \multicolumn{1}{l}{}                             & \multicolumn{1}{l}{}                                                                    &                   &                      &                      &                      &                      &                      &                      &                      &                      \\
                                 &                                            \multirow{3}{*}{6}                                & \multirow{3}{*}{0.8}                                                                    & mn-B              & 68.3                 & 0.143 (0.024)        &                      & 53.2                 & 0.139 (0.025)        &                      & 38.0                 & 0.132 (0.024)        \\
                                 &                                                                                              &                                                                                         & CB                & 92.8                 & 0.290 (0.087)        &                      & 92.8                 & 0.389 (0.117)        &                      & 93.8                 & 0.535 (0.162)        \\
                                 &                                                                                              &                                                                                         & MN-CB             & \textbf{93.8}        & 0.291 (0.088)        &                      & \textbf{94.0}        & 0.391 (0.118)        &                      & \textbf{94.0}        & 0.536 (0.163)        \bstrut \\  \hline \hline
\end{tabular}%
}
\caption{Results for simulation Scenario 2 (Small number of clusters). Estimates of 95\% confidence interval coverage and length for the coefficient of the $X_1 A_1$ interaction effect, $\psi_{11}$, estimated in the first-stage estimation for $N$ = 20 clusters with $n_i$ = 80 individuals per cluster. Largest coverage for each setting is shown in bold font. $\rho$ refers to the intra-cluster correlation (ICC) used to generate the simulation data; $p$ refers to the degree of non-regularity; Coverage (\%) represents the coverage of the 95\% confidence interval; Length represents the length of the 95\% confidence interval. Abbreviations: mn-B: $m$-out-of-$n$ Bootstrap; CB: Cluster Bootstrap; MN-CB: the proposed $M$-out-of-$N$ Cluster Bootstrap.}
\label{tab:tab2}
\end{table}

\textbf{Scenario 3 ($\bm{N = 30}$, $\bm{n_i \sim Unif(10,30)}$):} 
As shown in Table~\ref{tab:tab3}, in the regular examples, MN-CB generally achieves strong coverage---often near or above 95\%---although CB occasionally matches or slightly exceeds MN-CB coverage only in certain cells with smaller ICC ($\rho=0.05$ or $0.1$) and moderate stage~1 effects. This slight advantage of CB can again be explained by the ``cost'' associated with resampling fewer clusters under regular conditions. Meanwhile, MN-CB-w, which incorrectly assumes independence, still provides coverage around 93--94\%, only slightly lower than MN-CB. Notably, mn-B shows a more pronounced drop in coverage as either $\rho$ or the stage~1 effect size increases, reinforcing the importance of accounting for clustering.
In contrast, in the non-regular examples, MN-CB clearly outperforms both CB and mn-B, maintaining coverage consistently near or above 95\%, particularly as ICC and effect sizes grow. MN-CB-w remains competitive with coverage typically around 94\%, though correct specification (MN-CB) still provides a modest improvement. These findings confirm that MN-CB robustly handles both non-regularity and clustering, and even incorrect correlation specification (MN-CB-w) yields reliable coverage in moderately sized clusters.

\begin{table}
\centering
\resizebox{\columnwidth}{!}{%
\begin{tabular}{ccclcccccccc}
\hline  \hline
\multirow{2}{*}{\textbf{\begin{tabular}[c]{@{}c@{}}Regularity \\ Setting\end{tabular}}} & \multirow{2}{*}{\textbf{Ex. \#}} & \multirow{2}{*}{\textbf{\begin{tabular}[c]{@{}c@{}}Stage 1\\ Effect Size\end{tabular}}} & \multirow{2}{*}{} & \multicolumn{2}{c}{$\bm\rho$\textbf{ = 0.05}}       & \textbf{}            & \multicolumn{2}{c}{$\bm\rho$\textbf{ = 0.1}}        & \textbf{}            & \multicolumn{2}{c}{$\bm\rho$\textbf{ = 0.2}}        \\ \cline{5-6} \cline{8-9} \cline{11-12} 
                                 &                                                                                              &                                                                                         &                   & Coverage (\%)        & Length               &                      & Coverage (\%)        & Length               &                      & Coverage (\%)        & Length                \\  \hline \\
\multirow{11}{*}{\begin{tabular}[c]{@{}c@{}}Regular\\ ($p = 0$)\end{tabular}} & \multirow{3}{*}{1} & \multirow{3}{*}{0.2}                                                                                   & mn-B              & 91.6                 & 0.216 (0.033)        &                      & 81.2                 & 0.214 (0.032)        &                      & 68.0                 & 0.208 (0.033)        \\
                                 &                                                                                              &                                                                                         & CB                & \textbf{96.0}        & 0.284 (0.062)        &                      & 94.1                 & 0.345 (0.076)        &                      & 94.4                 & 0.445 (0.100)        \\
                                 &                                                                                              &                                                                                         & MN-CB-w           & 95.8                 & 0.289 (0.063)        &                      & 94.1                 & 0.350 (0.078)        &                      & 94.0                 & 0.453 (0.102)        \\
                                 &                                                                                              &                                                                                         & MN-CB             & 95.4                 & 0.288 (0.064)        &                      & \textbf{95.4}        & 0.350 (0.077)        &                      & \textbf{96.0}        & 0.451 (0.100)        \\
                                 & \multicolumn{1}{l}{}                                                                         & \multicolumn{1}{l}{}                                                                    &                   &                      &                      &                      &                      &                      &                      &                      &                      \\
                                 & \multirow{3}{*}{2}                                                                           & \multirow{3}{*}{0.5}                                                                    & mn-B              & 88.0                 & 0.215 (0.031)        &                      & 82.0                 & 0.213 (0.030)        &                      & 65.2                 & 0.207 (0.030)        \\
                                 &                                                                                              &                                                                                         & CB                & \textbf{95.2}        & 0.286 (0.066)        &                      & 94.8                 & 0.348 (0.081)        &                      & 94.0                 & 0.446 (0.103)        \\
                                 &                                                                                              &                                                                                         & MN-CB-w           & \textbf{95.2}        & 0.290 (0.067)        &                      & \textbf{95.2}        & 0.354 (0.081)        &                      & 94.1                 & 0.452 (0.104)        \\
                                 &                                                                                              &                                                                                         & MN-CB             & 94.8                 & 0.290 (0.066)        &                      & \textbf{95.2}        & 0.353 (0.082)        &                      & \textbf{96.0}        & 0.450 (0.103)        \\
                                 & \multicolumn{1}{l}{}                                                                         & \multicolumn{1}{l}{}                                                                    &                   &                      &                      &                      &                      &                      &                      &                      &                      \\
                                 & \multirow{3}{*}{3}                                                                           & \multirow{3}{*}{0.8}                                                                    & mn-B              & 88.4                 & 0.214 (0.028)        &                      & 79.9                 & 0.211 (0.029)        &                      & 64.3                 & 0.205 (0.028)        \\
                                 &                                                                                              &                                                                                         & CB                & 93.0                 & 0.281 (0.063)        &                      & 93.2                 & 0.341 (0.075)        &                      & 93.0                 & 0.437 (0.093)        \\
                                 &                                                                                              &                                                                                         & MN-CB-w           & 93.2                 & 0.286 (0.064)        &                      & 93.6                 & 0.348 (0.078)        &                      & 94.0                 & 0.445 (0.093)      \\
                                 &                                                                                              &                                                                                         & MN-CB             & \textbf{94.4}        & 0.288 (0.066)        &                      & \textbf{95.2}        & 0.349 (0.079)        &                      & \textbf{95.8}        & 0.447 (0.098)         \bstrut \\  \hline
\multicolumn{1}{l}{}             & \multicolumn{1}{l}{}                                                                         & \multicolumn{1}{l}{}                                                                    &                   & \multicolumn{1}{l}{} & \multicolumn{1}{l}{} & \multicolumn{1}{l}{} & \multicolumn{1}{l}{} & \multicolumn{1}{l}{} & \multicolumn{1}{l}{} & \multicolumn{1}{l}{} & \multicolumn{1}{l}{} \\
\multirow{11}{*}{\begin{tabular}[c]{@{}c@{}}Non-regular\\ ($p \neq 0$)\end{tabular}} & \multirow{3}{*}{4} & \multirow{3}{*}{0.2}                                                                    & mn-B              & 92.8                 & 0.250 (0.031)        &                      & 83.5                 & 0.246 (0.033)        &                      & 68.6                 & 0.212 (0.031)        \\
                                 &                                                                                              &                                                                                         & CB                & 93.3                 & 0.284 (0.062)        &                      & 94.1                 & 0.347 (0.076)        &                      & 91.7                 & 0.445 (0.100)        \\
                                 &                                                                                              &                                                                                         & MN-CB-w           & 94.9                 & 0.298 (0.065)        &                      & \textbf{94.8}        & 0.364 (0.079)        &                      & 93.3                 & 0.467 (0.104)        \\
                                 &                                                                                              &                                                                                         & MN-CB             & \textbf{95.0}        & 0.300 (0.065)        &                      & \textbf{94.8}        & 0.362 (0.080)        &                      & \textbf{94.6}        & 0.466 (0.102)        \\
                                 & \multicolumn{1}{l}{}                                                                         & \multicolumn{1}{l}{}                                                                    &                   &                      &                      &                      &                      &                      &                      &                      &                      \\
                                 & \multirow{3}{*}{5}                                                                           & \multirow{3}{*}{0.5}                                                                    & mn-B              & 88.2                 & 0.219 (0.029)        &                      & 79.5                 & 0.216 (0.029)        &                      & 63.5                 & 0.211 (0.030)        \\
                                 &                                                                                              &                                                                                         & CB                & 94.2                 & 0.287 (0.064)        &                      & 93.2                 & 0.348 (0.077)        &                      & 93.0                 & 0.450 (0.099)        \\
                                 &                                                                                              &                                                                                         & MN-CB-w           & \textbf{95.0}        & 0.292 (0.065)        &                      & 93.4                 & 0.354 (0.078)        &                      & 93.6                 & 0.459 (0.099)        \\
                                 &                                                                                              &                                                                                         & MN-CB             & \textbf{95.0}        & 0.291 (0.064)        &                      & \textbf{94.4}        & 0.355 (0.080)        &                      & \textbf{94.4}        & 0.456 (0.104)        \\
                                 & \multicolumn{1}{l}{}                                                                         & \multicolumn{1}{l}{}                                                                    &                   &                      &                      &                      &                      &                      &                      &                      &                      \\
                                 & \multirow{3}{*}{6}                                                                           & \multirow{3}{*}{0.8}                                                                    & mn-B              & 86.6                 & 0.218 (0.033)        &                      & 76.4                 & 0.215 (0.033)        &                      & 64.3                 & 0.209 (0.032)        \\
                                 &                                                                                              &                                                                                         & CB                & 92.7                 & 0.284 (0.059)        &                      & 92.9                 & 0.345 (0.071)        &                      & 92.5                 & 0.442 (0.092)        \\
                                 &                                                                                              &                                                                                         & MN-CB-w           & 93.5                 & 0.291 (0.061)        &                      & 93.1                 & 0.352 (0.072)        &                      & 93.1                 & 0.450 (0.094)        \\                
                                 &                                                                                              &                                                                                         & MN-CB             & \textbf{95.4}        & 0.289 (0.059)        &                      & \textbf{93.8}        & 0.353 (0.070)        &                      & \textbf{95.0}        & 0.455 (0.094)        \bstrut \\  \hline \hline
\end{tabular}
}
\caption{Results for simulation Scenario 3 (Moderate number of clusters and varied number of sample in each cluster). Estimates of 95\% confidence interval coverage and length for the coefficient of the $X_1 A_1$ interaction effect, $\psi_{11}$, estimated in the first-stage estimation for $N$ = 30 clusters with individuals per cluster $n_i \sim Unif(10,30)$. Largest coverage for each setting is shown in bold font. $\rho$ refers to the intra-cluster correlation (ICC) used to generate the simulation data; $p$ refers to the degree of non-regularity; Coverage (\%) represents the coverage of the 95\% confidence interval; Length represents the length of the 95\% confidence interval. Abbreviations: mn-B: $m$-out-of-$n$ Bootstrap; CB: Cluster Bootstrap; MN-CB-w: the proposed $M$-out-of-$N$ Cluster Bootstrap with wrong (independence) working correlation model; MN-CB: the proposed $M$-out-of-$N$ Cluster Bootstrap with correct (exchangeable) working correlation model.}
\label{tab:tab3}
\end{table}

\subsection{Aim 2: Impact of Working Correlation Models on Efficiency of MN-CB}
\label{simulation:aim2}

In this secondary simulation aim, we further explore the efficiency of the proposed MN-CB under a correctly specified working correlation model (exchangeable) compared to the wrongly specified one (independence). Performance is evaluated by bias, SE, and mean squared error (MSE) of the $\widehat \Psi_1$.
In comparing estimation efficiency between MN-CB and MN-CB-w (\ref{tab:tab4}), negligible differences in bias (all below 0.001) are observed, as both methods provide unbiased estimates (proof in Appendix 1.2.1). 
MN-CB consistently yields slightly lower SE and MSE, demonstrating an efficiency gain from correctly specifying the working correlation structure, especially as ICC increases (e.g., for Ex. 1, efficiency gains in MSE are 1.2\%, 4.0\%, and 10.1\% for $\rho = 0.05, 0.1,$ and $ 0.2$, respectively).
This result is theoretically justified because the correct model better captures within-cluster dependence. 
Although statistically meaningful, these observed differences likely have modest practical implications, as MN-CB-w maintains near-nominal coverage rates. Thus, correctly specifying the working correlation improves efficiency, particularly at higher ICC, though incorrect specification still yields robust results.

For exploratory purpose, we also designed three additional simulation examples representing ``near non-regular" conditions ($p=0$ and stage 2 effect size $\zeta \approx 0.2$) under three stage~1 effect sizes (0.2, 0.5, 0.8). Results are qualitatively similar to the main findings. Detailed tables and further discussion are provided in Web Appendix C for interested readers.

Overall, MN-CB provides stable coverage across regular and non-regular cases, various cluster counts, and ICC values. Although CB yields similar or better results in a few highly regular settings, such scenarios may rarely occur in real-world cSMART data. Therefore, MN-CB’s greater adaptability makes it a strong and reliable technique for examining moderators in cAI from cSMART data.

\begin{table}
\centering
\resizebox{\columnwidth}{!}{%
\begin{tabular}{ccclccccccccccc}
\hline \hline
\multirow{2}{*}{\textbf{\begin{tabular}[c]{@{}c@{}}Regularity\\ Setting\end{tabular}}} & \multirow{2}{*}{\textbf{Ex. \#}} & \multirow{2}{*}{\textbf{\begin{tabular}[c]{@{}c@{}}Stage 1\\ Effect Size\end{tabular}}} & \multirow{2}{*}{} & \multicolumn{3}{c}{$\bm\rho$\textbf{ = 0.05}} & & \multicolumn{3}{c}{$\bm\rho$\textbf{ = 0.1}} & & \multicolumn{3}{c}{$\bm\rho$\textbf{ = 0.2}} \\ \cline{5-7} \cline{9-11} \cline{13-15} 
& & & & \textbf{Bias} & \textbf{SE} & \textbf{MSE} & & \textbf{Bias} & \textbf{SE} & \textbf{MSE} & & \textbf{Bias} & \textbf{SE} & \textbf{MSE} \\ \hline
\multicolumn{15}{l}{} \\ 
\multirow{8}{*}{\begin{tabular}[c]{@{}c@{}}Regular\\ ($p=0$)\end{tabular}} & \multirow{2}{*}{1} & \multirow{2}{*}{0.2} & MN-CB-w & 0.000 & 0.064 & $4.15\times10^{-3}$ & & 0.001 & 0.080 & $6.40\times10^{-3}$ & & 0.003 & 0.103 & $1.06\times10^{-2}$ \\
& & & MN-CB & -0.001 & 0.064 & $4.10\times10^{-3}$ & & -0.001 & 0.078 & $6.15\times10^{-3}$ & & -0.004 & 0.098 & $9.54\times10^{-3}$ \\
& & & & & & & & & & & & & & \\ 
& \multirow{2}{*}{2} & \multirow{2}{*}{0.5} & MN-CB-w & 0.001 & 0.068 & $4.63\times10^{-3}$ & & 0.003 & 0.081 & $6.57\times10^{-3}$ & & -0.003 & 0.107 & $1.15\times10^{-2}$ \\
& & & MN-CB & 0.001 & 0.068 & $4.63\times10^{-3}$ & & 0.002 & 0.082 & $6.71\times10^{-3}$ & & 0.003 & 0.106 & $1.12\times10^{-2}$ \\
& & & & & & & & & & & & & & \\ 
& \multirow{2}{*}{3} & \multirow{2}{*}{0.8} & MN-CB-w & -0.003 & 0.069 & $4.77\times10^{-3}$ & & -0.003 & 0.084 & $7.07\times10^{-3}$ & & -0.005 & 0.112 & $1.26\times10^{-2}$ \\
& & & MN-CB & -0.002 & 0.066 & $4.31\times10^{-3}$ & & -0.003 & 0.082 & $6.76\times10^{-3}$ & & -0.004 & 0.105 & $1.10\times10^{-2}$ \bstrut \\ \hline
\multicolumn{15}{l}{} \\ 
\multirow{8}{*}{\begin{tabular}[c]{@{}c@{}}Non-regular\\ ($p\neq0$)\end{tabular}} & \multirow{2}{*}{4} & \multirow{2}{*}{0.2} & MN-CB-w & 0.001 & 0.071 & $5.04\times10^{-3}$ & & 0.001 & 0.086 & $7.40\times10^{-3}$ & & -0.002 & 0.111 & $1.23\times10^{-2}$ \\
& & & MN-CB & -0.002 & 0.071 & $4.99\times10^{-3}$ & & 0.001 & 0.085 & $7.18\times10^{-3}$ & & 0.002 & 0.109 & $1.20\times10^{-2}$ \\
& & & & & & & & & & & & & & \\ 
& \multirow{2}{*}{5} & \multirow{2}{*}{0.5} & MN-CB-w & 0.006 & 0.069 & $4.80\times10^{-3}$ & & 0.007 & 0.084 & $7.10\times10^{-3}$ & & 0.009 & 0.114 & $1.31\times10^{-2}$ \\
& & & MN-CB & 0.006 & 0.068 & $4.66\times10^{-3}$ & & 0.007 & 0.083 & $6.86\times10^{-3}$ & & 0.008 & 0.107 & $1.15\times10^{-2}$ \\
& & & & & & & & & & & & & & \\ 
& \multirow{2}{*}{6} & \multirow{2}{*}{0.8} & MN-CB-w & 0.000 & 0.070 & $4.90\times10^{-3}$ & & 0.002 & 0.087 & $7.57\times10^{-3}$ & & 0.000 & 0.110 & $1.21\times10^{-2}$ \\
& & & MN-CB & 0.000 & 0.069 & $4.69\times10^{-3}$ & & 0.000 & 0.085 & $7.17\times10^{-3}$ & & 0.003 & 0.107 & $1.15\times10^{-2}$ \bstrut \\ \hline \hline
\end{tabular}
}
\caption{Results for simulation aim 2. Estimates of bias, standard error (SE), mean squared error (MSE) for the coefficient of the $X_1 A_1$ interaction effect, $\psi_{11}$, estimated in the first-stage estimation for $N$ = 30 clusters with $n_i \sim Unif(10,30)$ individuals per cluster. Largest coverage for each setting is shown in bold font. $\rho$ refers to the intra-cluster correlation (ICC) used to generate the simulation data; $p$ refers to the degree of non-regularity. Abbreviations: MN-CB-w: the proposed $M$-out-of-$N$ Cluster Bootstrap with wrong (independence) working correlation model; MN-CB: the proposed $M$-out-of-$N$ Cluster Bootstrap with correct (exchangeable) working correlation model.}
\label{tab:tab4}
\end{table}

\section{Application to ADEPT Data}\label{S:example-using-ADEPT}


The Adaptive Implementation of Effective Programs Trial (ADEPT) is a cSMART trial with type III design. 
It was conducted at community-based, outpatient clinics in Michigan and Colorado and was designed to determine how best to support nonresponsive clinics in implementing evidence-based practices (EBPs), which have been shown to improve patient-level outcomes for individuals with anxiety or depression, post-traumatic stress disorder, autism, and others \cite{kilbourne2014protocol}. However, due to a large degree of heterogeneity across clinics, we would expect that clinics may respond differently to varying levels of implementation support.
One of the stated objectives of this trial was to identify clinic-level factors at each intervention stage, if any, that could be used to tailor the level of intervention necessary to ensure the clinic would successfully implement EBPs across their practice. At the outset of the study, all participating clinics were offered training in replicating effective programs (REP), a system designed to help them implement EBPs. 
\begin{figure}
    \centering
    \resizebox{\textwidth}{!}{%
    \includegraphics{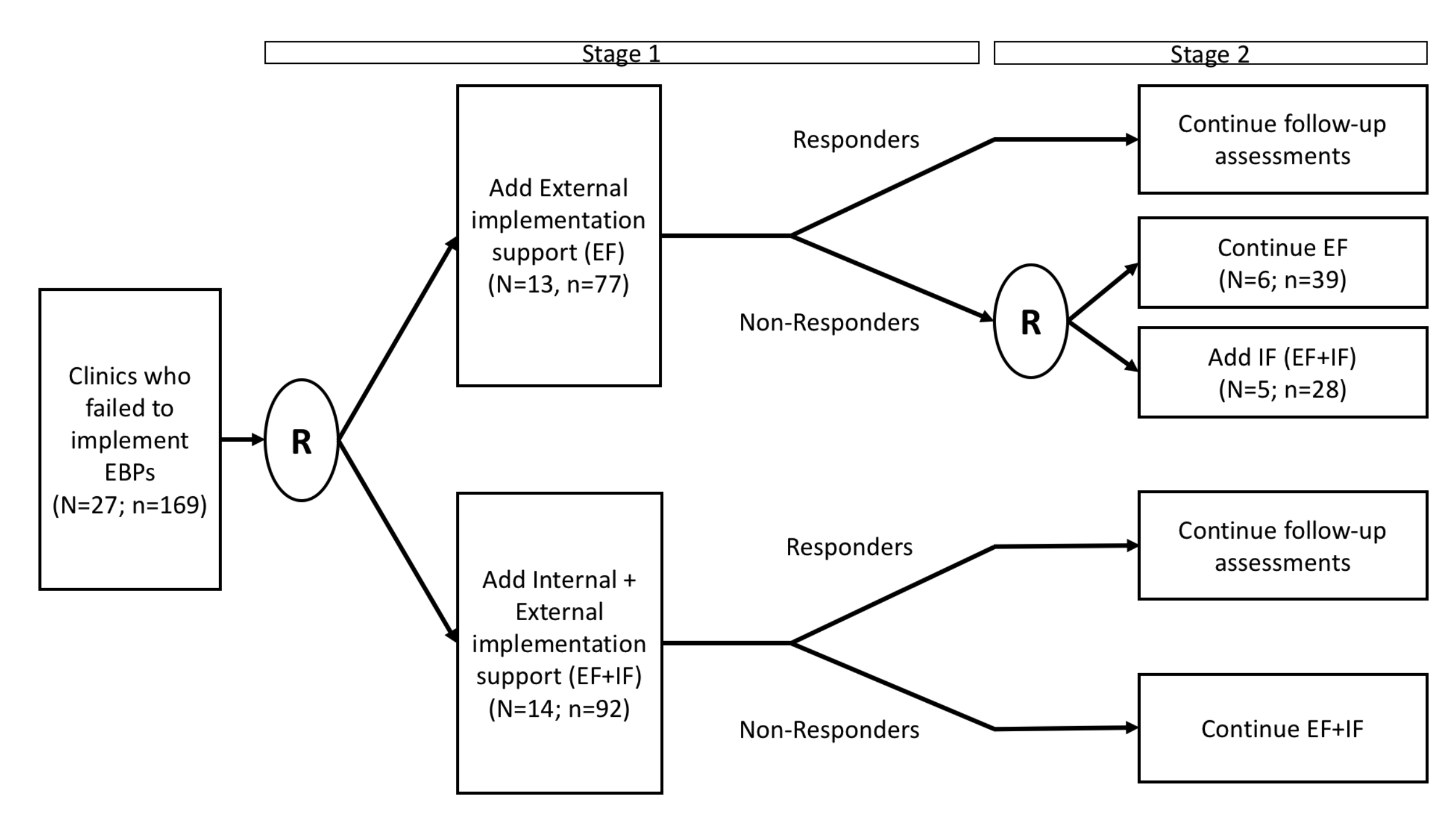}
    }
    \caption{Clustered Sequential Multiple Assignment Randomized Trial (cSMART) designed to evaluate the use of Internal (IF) and/or External (EF) implementation support for primary and mental health clinics who failed to implement evidence-based practices (EBPs) after a 6 month run-in period. R indicates 1:1 randomization performed. $N$ = number of clinics; $n$ = number of patients within the $N$ clinics. 
    }
\label{fig:fig1}
\end{figure}
As shown in Figure~\ref{fig:fig1}, the first randomization event included only those clinics that failed to effectively implement the EBPs. These clinics were randomized 1:1 to receive one of two different intervention support systems: external support alone (EF) or both external and internal support (EF+IF). Refer to \cite{kilbourne2014protocol} and \cite{smith2019change} for additional information about REP and the interventions EF and IF. Following the stage 1 response assessment, which occurred 6 months after the first randomization, clinics that received EF at stage 1 were withdrawn from EF if they had effectively implemented the EBPs. On the other hand, EF-clinics who had not been successful in implementing EBPs were re-randomized to either continue EF or to add internal support (i.e., EF+IF). All clinics who received EF+IF at stage 1 either stopped (or continued) EF+IF if they had (or had not) effectively implemented the EBPs. Patient-level outcomes were collected at baseline and after completing the first and second treatment stages.

Thirteen clinics were randomized to receive EF and 14 clinics to EF+IF at the first stage. 
Of the clinics randomized to EF at stage 1, six clinics were re-randomized at stage 2 to continue EF alone and five clinics were re-randomized to add the IF support. The intraclass correlation of outcomes within each clinic was estimated to be 0.23. Refer also to \cite{smith2019change} for summary statistics describing the patient cohort and results of the primary analysis.

We apply our proposed clustered $Q$-learning with 
MN-CB
to evaluate whether clinic-level factors can be used to tailor EBP-implementation support at stage 1 or stage 2,
aiming to improve mental health outcomes for patients with mood disorders.
The primary outcome is mental health quality of life (MHQOL), assessed for each patient following the stage 2 intervention. MHQOL is measured using the Short Form-12 (SF-12) \cite[][]{ware199612}, which is scored from $0 - 100$ with higher scores indicating better MHQOL. Interventions at each stage, denoted $A_1$ and $A_2$, respectively, include EF+IF (versus EF alone), both stages coded as 1 (or $-1$). Cluster-level covariates collected prior to the first randomization include whether clinics are rural or urban ($1/-1$), located in Michigan or Colorado ($1/-1$), and have a higher or lower than average site-aggregated MHQOL stratum ($1/-1$). 
Additionally, a factor coded as 1 $(-1)$ represents a higher (or lower) site-aggregated mean MHQOL level. At stage 1 we consider two candidate tailoring variables: the site-aggregated mean MHQOL level preceding the first randomization (M6-MH) and the state in which the clinic is located; high MHQOL strata and the state of Michigan (MI) were used as reference categories. At stage 2, we evaluate site-aggregated mean MHQOL level immediately preceding the second randomization (M12-MH). Variables used to stratify randomization are also included in both stage-specific regression models.

We perform this analysis using the implementation guidelines described in Web Appendix A. The stage 2 $Q$-function model is specified as follows: 
\vspace{-3pt}
\begin{equation}
    Q_2\left(\bm{H}_2, A_2\right)=\gamma_{20}+\gamma_{21} \mathrm{Rural} + \beta_{21}\text{M12-MH}+\left\{\psi_{20}+\psi_{21}\text{M12-MH}\right\}  \times A_2.
\end{equation}
\vspace{-35pt}

For patients treated at clinics re-randomized at stage 2, the stage 1 pseudo-outcome is:
\vspace{-3pt}
\begin{equation}
\widetilde{Y}_{M H,ij}=\widehat{\gamma}_{20}+\widehat{\gamma}_{21}\mathrm{Rural}_i+ \widehat{\beta}_{21}\text{M12-MH}_i+\left|\widehat{\psi}_{20}+\widehat{\psi}_{21}\text{M12-MH}_i\right|.
\end{equation}
\vspace{-35pt}

The stage 1 $Q$-function is modeled by: 
\vspace{-3pt}
\begin{equation}
    Q_1\left({\bm{H}_1}, A_1\right)= 
    \gamma_{10}+\gamma_{11} \mathrm{Rural} +\beta_{11}\mathrm{MI}+\beta_{12}\text{M6-MH}
    +\left\{\psi_{10}+\psi_{11}\mathrm{MI}+\psi_{12}\text{M6-MH}\right\} \times A_1.
\end{equation}
\vspace{-35pt}

We perform the MN-CB resampling at both stages with $B=2500$ iterations and remove any bootstrap resample that fails to generate estimates due to singularity. Given the exploratory nature of this analysis, we estimate confidence intervals based on a pre-specified significance level of $\alpha=0.10$. Due to the high degree of missingness in the overall MHQOL for patients treated at clinics that were re-randomized at stage 2, as well as the composition of the sites re-randomized at stage 2 (i.e., the absence of urban clinics), we utilize multiply imputed datasets with appropriate combining rules \cite[][]{little2019statistical}.

Estimated regression coefficients and associated 
confidence intervals 
constructed by MN-CB
are shown in Table~\ref{tab:tab5}. To determine whether the set of candidate variables may be useful in tailoring a cAI to optimize individual-level counterfactual outcomes across the population of interest, we examine the interaction effects of cluster-level covariates with the intervention EF+IF at both stages (Rows 5, 6, 10 in Table~\ref{tab:tab5}). At stage 1, the estimated 90\% confidence intervals for the stage 1 EF+IF interventions with state and high clinic mean Month 6 MHQOL are ($-3.11, 2.89$) and ($-1.44, 3.24$), respectively, both of which include zero. Similarly, at stage 2, the estimated 90\% confidence interval for the EF+IF interaction with high mean Month 12 MHQOL is ($-3.34, 2.25$), suggesting that there is insufficient evidence in our data to conclude that any of these candidate variables would be useful in further tailoring a cAI to support the implementation of EBPs. As this was an exploratory analysis and the study was not powered based on this statistical objective, it is possible that either (i) these tailoring variables should not be used to further refine clinic-level interventions to improve the implementation of EBPs at primary care and mental health clinics located in Michigan and Colorado; or (ii) there is insufficient power in our dataset to identify these effects.

\begin{table}
\centering
\resizebox{0.7\columnwidth}{!}{%
\begin{tabular}{llcc}
\hline \hline
\multicolumn{2}{l}{\textbf{}} & \textbf{Estimate} & \textbf{90\% CI} \\ \hline
\multicolumn{2}{l}{\textbf{Stage 1 Variables}} & \multicolumn{1}{l}{} & \multicolumn{1}{l}{} \\
 & Rural (vs. Not Rural) & -6.80 & (-15.20, -7.00) \\
 & Michigan (vs. Colorado) & 0.64 & (-2.65, 3.16) \\
 & High Mean M6 MHQOL (vs. Low) & 0.40 & (-2.11, 2.42) \\
 & EF+IF (vs. EF alone) & -3.74 & (-4.34, 0.97) \\
 & (EF+IF):(Michigan) & 1.31 & (-3.11, 2.89) \\
 & (EF+IF):(High Mean M6 MHQOL) & 0.82 & (-1.44, 3.24) \\ \hline
\multicolumn{2}{l}{\textbf{Stage 2 Variables}} &  &  \\ 
 & Rural (vs. Not Rural) & -5.94 & (-14.80, 1.20) \\
 & High Mean M12 MHQOL (vs. Low) & -1.53 & (-3.90, 0.82) \\
 & EF+IF (vs. EF alone) & -0.19 & (-2.65, 2.30) \\
 & (EF+IF):(High Mean M12 MHQOL) & -0.60 & (-3.34, 2.25) \\ \hline\hline
\end{tabular}}
\caption{Estimated stage 1 and stage 2 regression coefficients and associated $90 \%$ $M$-out-of-$N$ Cluster Bootstrap confidence intervals (CI). The outcome of interest is patient-level Month 18 Mental Health Quality of Life (MHQOL). M6: Month 6 (prior to first randomization); M12: Month 12 (prior to second randomization); Interventions at both stage 1 and stage 2 include EF+IF (external and internal implementation support) versus EF alone. All covariates are measured at the cluster level.}
\label{tab:tab5}
\end{table}

\section{Discussion}

In this manuscript, we propose a clustered $Q$-learning approach with the $M$-out-of-$N$ Cluster Bootstrap to make statistical inference on whether pre-specified cluster-level candidate tailoring variables may be useful to further adapt multi-stage interventions for improving individual-level outcomes. This work addresses an essential methodological gap for cSMARTs, which often request to justify the use of these tailoring variables in order to construct or refine interventions at the cluster level.
Our simulation findings indicate that, with a moderate to large number of clusters ($N \ge 30$) and a reasonable number of individuals per cluster, parameter estimates and confidence intervals exhibit low bias and coverage near or slightly above the nominal level across both regular and non-regular data-generating settings. Even with fewer clusters that we examine as low to $N = 20$, $M$-out-of-$N$ bootstrapping remains competitive, especially when the first-stage effect size is moderate or large. In simple, highly regular scenarios, the standard Cluster Bootstrap can sometimes match or slightly surpass $M$-out-of-$N$ coverage, but real-world cSMARTs typically involve more complex conditions where our method shows distinct advantages. Overall, the clustered $Q$-learning framework with $M$-out-of-$N$ Cluster Bootstrap appears flexible, robust, and well suited to evaluate multi-stage tailoring effects in cSMART designs.

We note three straightforward extensions of clustered $Q$-learning. First, while we employ a parametric regression framework to model the multi-stage clustered $Q$-functions, alternative or nonparametric strategies may be suitable when sample sizes are large or when effect estimation is secondary. 
Second, our primary focus is on a continuous outcome; however, the generalized estimating equations framework readily permits extension to binary, count, or other outcomes using the appropriate link function. 
Third, although this work is motivated by cSMART data, our methods could be extended to observational multi-stage studies with clustering, where propensity-adjustment techniques or more flexible modeling frameworks \cite[][]{moodie2012q, chakraborty2013inference} might be necessary. 

Two main limitations should be acknowledged. First, we adopt a relatively simple data-generating mechanism for an unrestricted cSMART with one binary baseline covariate and one binary intermediate covariate. Although designed to represent varying degrees of (non-)regularity, this simplified structure may not fully capture the complexities of real-world cSMARTs. For instance, we considered only cluster-level covariates as effect modifiers, although individual-level covariates can easily be included.
Additionally, we rely on parametric modeling assumptions at each stage of the Q-learning procedure, which may lead to bias if the assumed models are misspecified in practice. Nonparametric or semiparametric extensions may offer greater flexibility, albeit with increased computational cost and complexity.

Despite these limitations, the cSMART design has grown increasingly popular over the last decade, with numerous studies employing cluster-level randomization and multi-stage decisions in fields ranging from mental health to organizational interventions \cite[e.g.,][]{fernandez2020quitsmart,kilbourne2018adaptive,quanbeck2020balanced}. Our proposed clustered $Q$-learning framework with the $M$-out-of-$N$ Cluster Bootstrap offers a practical, conceptually clear approach to identifying key tailoring variables and supporting robust inferences—precisely the goal cSMARTs were designed to achieve. By addressing the critical question of whether cluster-level tailoring variables can reliably inform multi-stage decision rules, this method stands to advance precision healthcare across a range of domains.

\section*{Acknowledgements}
This work was supported by grants from the National Institutes of Health (P50DA054039 and R01DA039901 to DA; R01MH114203 and R01MH099898 to AMK; RO1DA047279 to AQ) and the Institute of Education Sciences (R324B220001 to DA). YS, KS are equal-effort lead authors, listed alphabetically by last name; AMK is the ADEPT study PI; and DA, LW are equal-effort senior authors. The content of this paper is solely the responsibility of the authors and does not necessarily represent the official views of the National Institutes of Health, the Institute of Education Science or the US Department of Veteran Affairs.

\section*{Data Availability}
The ADEPT study data used to illustrate the methods in this paper are available upon request to, and approval by, AMK. The data are not publicly available. 




\providecommand{\newblock}{}

\end{document}